\documentstyle[aps,prl,multicol,amssymb,epsf,float,psfig]{revtex}

\newcommand{\ket}[1]{\left | #1 \right \rangle}
\newcommand{\bra}[1]{\left \langle #1 \right |}

\title{Efficient factorization with a single pure qubit and $log N$ mixed qubits}
\author{S. Parker and M.B. Plenio}
\address{Optics Section, The Blackett Laboratory, Imperial College,
London SW7 2BW, England}
\date{\today}

\begin{document}
\draft
\maketitle
\begin{abstract}
It is commonly assumed that Shor's quantum algorithm
for the efficient factorization of a large number $N$ requires a pure initial
state. Here we demonstrate that a single pure qubit together with a collection
of $log_2 N$ qubits in an arbitrary mixed state is sufficient to implement Shor's
factorization algorithm efficiently.
\end{abstract}


\begin{multicols}{2}

The discovery of a quantum algorithm for the efficient
factorization of large numbers \cite{Shor95} has started a rapid
development of quantum information processing \cite{Summary}.
Following this ground-breaking result a number of experimentally
realizable proposals for the implementation of quantum computers
have been made, for example, in ion trap systems \cite{Cirac} or
Nuclear Magnetic Resonance (NMR) schemes \cite{NMR}. These systems
are distinguished by a low decoherence rate combined with a
comparatively high gate speed and therefore promise the
possibility of executing many quantum gates. While noise in these
systems can be made small in principle, it nevertheless imposes
limitations to the maximal size of the computation \cite{noise}
and to the achievable quality (e.g. the purity) of the
initial state of the quantum computer. It would therefore be
interesting to see whether a quantum computation necessarily
requires the preparation of an initial state of high purity, or
whether some parts of the quantum computer may be left in a mixed
state. Such a result would be of particular interest in NMR
systems in which it is difficult to prepare physically pure
quantum states of nuclear spins.

The use of mixed states in quantum algorithms has had little discussion as yet.
Note, however, the work of Schulman and Vazirani \cite{Vazirani} in which they
demonstrated that, starting from a set of qubits each in a thermal state, one
can obtain a certain number of pure qubits using a quantum algorithm. These
were then envisaged to be used for a quantum computation, while all the other
qubits which are in a mixed state are discarded. If the initial
states are in a thermal mixture at high temperature, the number of mixed quantum
states and quantum gates required to obtain even a single pure qubit is very high. It
would greatly enhance the efficiency of this approach if it would be possible to
reduce the necessary number of pure qubits as much as possible at the expense of
employing some of the mixed qubits in the actual quantum computation. Recently,
Knill and Laflamme \cite{Knill} have investigated the power of quantum computation
when only a single pure qubit together with a supply of {\em maximally} mixed
states is available. They were able to construct a problem that such a system
can solve more efficiently than the best currently {\em known} classical algorithm.

It would be interesting to see whether these ideas can be extended to
other problems of practical relevance. In this paper we demonstrate that
a {\em single} pure qubit together with an initial supply of $\log_2 N$
qubits in an arbitrarily mixed state is sufficient to implement Shor's
algorithm for the factorization of the number $N$ efficiently. This is
the smallest number of pure states that can achieve this task. We also
demonstrate that the efficiency of the modified algorithm is essentially
independent of the degree of mixing of the $\log_2 N$ qubits.

We proceed by outlining the problem addressed in Shor's algorithm, followed
by the formulation of Shor's algorithm introduced in \cite{ek}.
Then we will describe the necessary modifications to this algorithm, that
will allow it to be executed using a single pure qubit and $\log_2 N$
qubits in a maximally mixed state.

The basis of Shor's algorithm is a classical order finding method
which, recast as a quantum algorithm, can be executed in polynomial time,
requiring only a polynomial amount of additional classical computation
to compute the factors of $N$. The factors of a number $N=pq$
can, with high probability, be found if the period or {\em order}, $r$,
(the lowest positive integer $x\neq 0$ such that $f_a(x) = 1$ ) of the
element $a$ in the space of the function
$
 f_a(x) = a^{x} \hbox{mod} \, N,
$
is known. Then, provided $a$ is coprime to $N$ (which can be checked classically
in polynomial time using Euclid's algorithm), there is a high probability that
$\hbox{gcd}(a^{\frac{r}{2}} \pm 1, N)$ yields a factor of $N$, where
$\hbox{gcd}(\alpha, \beta)$ denotes the greatest common divisor of $\alpha$ and
$\beta$ which, again, can be determined efficiently using Euclid's algorithm \cite{Shor95}.

We begin by examining the formulation of Shor's algorithm as given in
\cite{ek} and use it as a basis to demonstrate the main result of this paper.
First of all we introduce the transformation
$U_a \ket{x} = \ket{ax \, \hbox{mod} \, N}$ where $x = 0,\cdots,N-1.$
Provided $a$ is coprime to $N$ this is a unitary transformation and has eigenvectors
\begin{equation}
\label{psis}
\ket{\psi_j} = \sum^{r-1}_{k=0} e^\frac{-2\pi i j k}{r} \ket{a^k \, \hbox{mod} \, N} \qquad
j = 0, \cdots, r-1
\end{equation}
with corresponding eigenvalues $e^\frac{2\pi i j}{r}$. Given one of these eigenvectors
we can apply $U_a$ to it and the value of $r$ will be encoded in the phase,
$e^\frac{2\pi i j}{r}$. This, however, is a global phase which we cannot
measure so instead we can use the "phase-kickback" technique \cite{ek} requiring
the conditional unitary transformation given by
\begin{eqnarray}
\label{uc}
    cU_a \ket{0} \ket{x} &=& \ket{0} \ket{x} \; ; \;
    cU_a \ket{1} \ket{x} = \ket{1} \ket{ax \,\hbox{mod} \,N}.
\end{eqnarray}
The effect of applying the controlled unitary transform to the state $(\ket{0} +
\ket{1})\ket{\psi_j}$ is
\begin{equation}
    \label{applyU}
    cU_a  (\ket{0} + \ket{1})\ket{\psi_j} = (\ket{0} + e^\frac{2\pi i
    j}{r}\ket{1})\ket{\psi_j}
\end{equation}
'kicking' the 'global' phase shift acquired on the second qubit into a relative
phase in the first qubit.  We can now perform measurements on the first qubit
which will allow us to estimate $r$, however, we cannot create the eigenstates
of $U_a$ without knowledge of $r$. Instead one can use the fact
\cite{ek}
that
$
%\begin{equation}
%\label{psiprop}
    \sum^{r-1}_{j=0} \ket{\psi_j} = \ket{1}
%\end{equation}
$
and conditionally apply $U_a$ to the state $\ket{1}$ (which obviously requires no
knowledge of $r$) in the second qubit
\begin{equation}
    \label{applyUto1}
    cU_a  (\ket{0} + \ket{1})\ket{1} = \sum^{r-1}_{j=0} (\ket{0} + e^\frac{2\pi i
    j}{r}\ket{1})\ket{\psi_j}.
\end{equation}
This state is, of course, entangled, so when we make measurements on the first qubit
we will get an estimate of $e^\frac{2\pi i j}{r}$, with $j$ (which corresponds to an
eigenstate) selected at random.

How do we estimate this phase and the value of $r$ accurately? The network in Fig.\@
\ref{Figure1} will give us, with a sufficient probability, the best $L$-bit estimate
of the value of $2^L j/r$ \cite{ek}.

\end{multicols}
\begin{minipage}{6.54truein}
 \begin{figure}[H]
 \begin{center}
  \leavevmode
  \epsfxsize=6.20truein
  \epsfbox{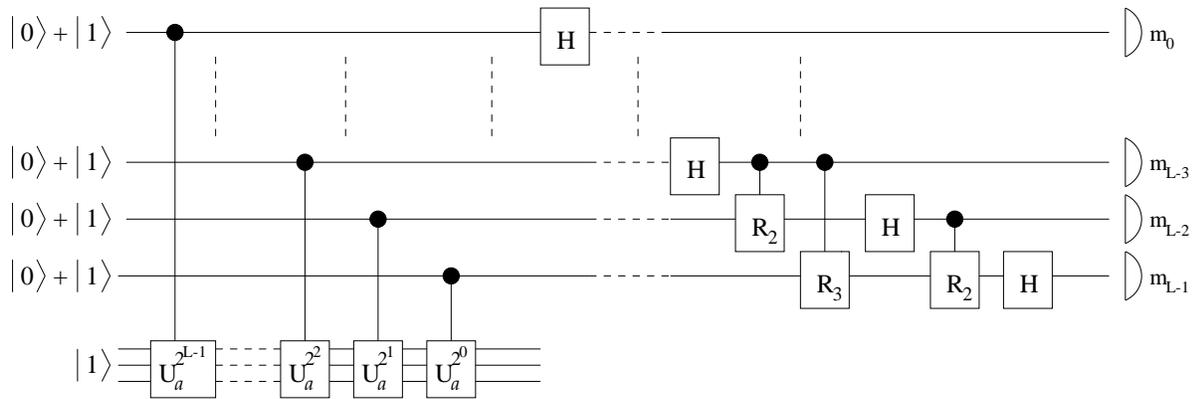}
 \end{center}
 \caption{An implementation of Shor's algorithm \protect\cite{ek}. The controlled $U_a$ operations produce phase shifts related to the order of $U_a$ and the remaining Hadamard transformations (H) and controlled rotations $R_j = {1 \, 0 \choose 0 \, \phi_j }$ with $\phi_j = e^{-2 \pi i/2^j}$ implement the inverse Fourier transform.}
 \label{Figure1}
 \end{figure}
\end{minipage}
\begin{multicols}{2}

As the algorithm proceeds it uses the controlled
$U_a, U_a^2, U_a^{2^2}, \cdots,
U_a^{2^L}$ transformations to
produce the 'kicked' phases $e^\frac{2\pi i j}{r}, e^\frac{2^2\pi i j}{r},
e^\frac{2^{3}\pi i j}{r},\cdots, e^\frac{2^{L-1}\pi
i j}{r}$ into the upper 'control' qubits.
The remaining operations on the control
qubits realise the quantum inverse Fourier transform. A measurement on each of these
qubits
produces a binary number $c = \sum_{i=0}^{L-1} 2^i m_i$ such that with a finite
probability $c/2^L$ is the best estimate of $j/r$ for some integer $j$ again selected
at random on measurement.

The first modification to this algorithm comes when we notice that
the gates within the Fourier transform are applied sequentially on
the qubits. Thus instead of performing the entire transform and
then making measurements on all control qubits afterwards we may
apply the single qubit (Hadamard) operation to the first qubit and
then measure it. The operations (controlled phase shifts)
controlled by this first qubit are then replaced by single qubit
operations {\em given the result of the measurement on the first}.
This 'semi-classical' modification \cite{griff} preserves the
probabilities of all measurement results.

Taking this further we need only insist on one control qubit and
the remaining $\lceil \log_2{N} \rceil$ qubits as we can 'recycle'
the control qubit after each measurement (Fig. \ref{Figure2}): we
perform all the necessary operations of the first control qubit
including measurements, followed by all the operations of the
second control qubit {\em on the same physical qubit system} given
the results of previous measurements, and so on \cite{Mosca E}.

We can, therefore, already implement Shor's algorithm with
$1+\lceil \log_2{N} \rceil$ pure qubits that is, one control qubit
and $\lceil \log_2{N} \rceil$ of the remaining qubits. We will
find later that we can also replace the $\lceil \log_2{N} \rceil$
pure qubits with $\lceil \log_2{N} \rceil$ maximally mixed qubits
and find the order $r$ efficiently (see also \cite{Mosca}). To see
why this is the case we first need to examine the unitary
transformation $U_a$ more closely.
\end{multicols}
\begin{minipage}{6.54truein}
 \begin{figure}[H]
 \begin{center}
  \leavevmode
  \epsfxsize=6.50truein
  \epsfbox{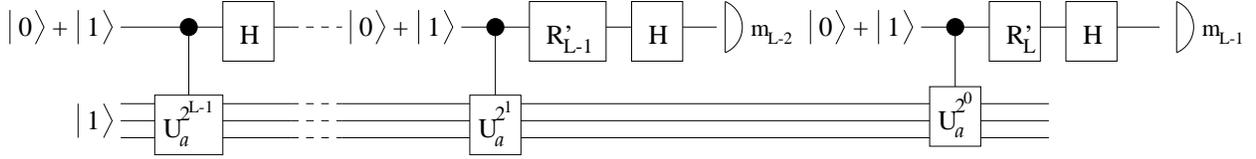}
 \end{center}
 \caption{An implementation of Shor's algorithm using only one control qubit which is recycled. $R'_j$ are now combinations of the rotations $R_j$ given the results of previous measurements: $R'_j = {1 \, 0 \choose 0 \, \phi'_j }$ with $\phi'_j = e^{-2 \pi i \sum^j_{k=2} m_{j-k}/2^k}$.}
 \label{Figure2}
 \end{figure}
\end{minipage}
\begin{multicols}{2}
The unitarity of the transform together with the fact that it maps a 'number' state
$\ket{x}$ to a
'number' state $\ket{ax \, \hbox{mod} \, N}$ means that on repeated application of
$U_a$ periodic sequences are induced on all the
numbers $x = 0,1,\cdots,N-1$, that is, there is an $R(x)$ such that $U_a^{R(x)} \ket{x}
= \ket{x}$. We may write the members of all possible sequences as
$
%\begin{equation}
%\label{writecyc}
\ket{g a^x \, \hbox{mod} \, N} \hbox{for some $g$ and $x$.}
%\end{equation}
$
For example, for $a=2$ and $N = 15$ on repeated application of
$U_a$ the possible sequences are
\begin{eqnarray}
\label{sequenceeg}
g=1: \qquad &&\ket{1} \rightarrow \ket{2} \rightarrow \ket{4} \rightarrow \ket{8}
\rightarrow \ket{1} \nonumber\\
g=3: \qquad &&\ket{3} \rightarrow \ket{6} \rightarrow \ket{12} \rightarrow \ket{9}
\rightarrow \ket{3} \nonumber \\
g=5: \qquad &&\ket{5} \rightarrow \ket{10} \rightarrow \ket{5} \nonumber \\
g=7: \qquad &&\ket{7} \rightarrow \ket{14} \rightarrow \ket{13} \rightarrow \ket{11}
\rightarrow \ket{7} .
% \\ g=0: \qquad &&\ket{0} \rightarrow \ket{0}
\end{eqnarray}

It is the first of these sequences (with $g=1$) whose number of members is what we
previously called the 'order', $r$, of $a$ modulo $N$ and it is this period  that we
need to find to factorize $N$. However, there is a relationship between the order of
the sequence with $g=1$ and the orders of all the other sequences with $g\ne 1$.  We will
label each of the different sequences by $d$ and the number of members in each sequence
by
$r_d$. $U_a$ obeys the condition $U_a^r = I$ so it is clear that $r_d|r$, that is, the
orders of all the sequences divides that of the sequence with $g=1$. In fact we will find
that
nearly all of the numbers $0,1,\cdots,N-1$ are contained within a sequence that has the
{\em same} order as the first sequence.
We can find a lower bound on the probability that for a number $g \in 0,1,\cdots,N-1$
the state $|ga^x \hbox{mod} N\rangle$ is contained within a sequence of order
$r$.\\

{\bf Theorem 1} {\it
Given two prime numbers $p$ and $q$ we define r as the lowest positive integer x such that $a^x
-1 \equiv 0 \, \hbox{mod} \, (p q)$ for an arbitrary integer $a$. Then $ga^x -g \equiv 0
\, \hbox{mod} \, (p q)$ with $x<r$ for at most $p+q-1$ values of $g$ in the interval $0
\le a \le pq-1$.}\\

{\bf Proof:} If $\hbox{gcd} (g,pq) = 1$ then $g(a^x-1) \equiv 0 \, \hbox{mod} \, p q
\Rightarrow a^x-1 \equiv 0 \, \hbox{mod} \, p q$ and therefore $x=r$. There are
$(p-1)(q-1)$ positive integers less than and coprime to $pq$, which proves the
theorem ${}_{\Box}$

We can now see that the probability, $P_r$, of picking $g$ at random such that the
lowest $x$ for which $ga^x \equiv g \, \hbox{mod} \, (p q)$ is $r$, is
$
%\begin{equation}
%\label{res}
P_r \ge (pq -(p+q-1))/pq = (p-1)(q-1)/pq
%\end{equation}
$
which approaches unity as $p$ and $q$ become large.

This tells us that if we set up an algorithm that actually finds the order of a random
sequence we still have a good chance that this order is in fact $r$.

The $r$ eigenstates of $U_a$ in equation \ref{psis} are orthogonal superpositions of the
members of the sequence with $g=1$. In exactly the same way we can form the remaining
$N-r$
eigenstates of $U_a$ as orthogonal superpositions of members of each of the other
sequences.
We write these as
\begin{equation}
\label{Npsis}
\ket{\psi^d_{j_d}} = \sum^{r_d -1}_{k = 0} e^{\frac{-2\pi i j_d k}{r_d}} \ket{g_d a^k \,
\hbox{mod} \, N}
\end{equation}
where $d$ labels the sequence and $j_d = 0, \cdots, r_d -1$ the eigenstates of $U_a$
within
the sequence $d$. $\ket{g_d}$ is the lowest member of the $d$th sequence. Each eigenstate has
corresponding
eigenvalue $e^{2 \pi i j_d/r_d}$ so using the same phase estimation techniques allows us
to estimate $j_d/r_d$ given the state $\ket{\psi^d_{j_d}}$. Again, this requires
knowledge of the sequences induced by $U_a$ so instead we may perform the phase
estimation
technique on the maximally mixed state
\begin{equation}
\label{maxmixis}
 \frac{{\bf 1}}{N} = \frac{1}{N} \sum^{N-1}_{k=0} \ket{k}\bra{k} =
\frac{1}{N} \sum_{d} \sum^{r_d-1}_{j_d=0} \ket{\psi^d_{j_d}}\bra{\psi^d_{j_d}}.
\end{equation}
Phase estimation now estimates the value of $j_d/r_d$ for $j_d$ and $d$ chosen at random
but as we have seen above nearly all the orders $r_d$ are equal to $r$.

Note that in Shor's original algorithm the $\lceil \log_2{N} \rceil$ qubits encode
a phase change into the control qubits which is quantum mechanically correlated to
eigenstates of $U_a$  our modification
encodes a phase change which is {\em classically} correlated
to the eigenstates. This includes not only the group of eigenstates consisting of
superpositions of elements in the first sequence (see Eq. \ref{sequenceeg})
but groups of eigenstates consisting
of superpositions of elements in each of the other sequences. However by theorem
1 most of these sequences have the same order and will encode the value
$r_d = r$ into the control qubits. This makes it intuitively clear that the algorithm
is still efficient. Note however that although the $\lceil \log_2{N} \rceil$ mixed
qubits are only classically correlated to the pure qubit, entanglement still exists
in the system:
one can partition the system into two halves one containing some mixed qubits and the
other containing the remaining mixed qubits and the pure qubit. Then it can be
checked, that this bipartite system can have negative partial transpose and is
therefore entangled \cite{NPT}.

In the following we will prove strictly that this modified version of Shor's
algorithm is indeed still efficient for order finding. Shor's algorithm requires
$O\left(\log \log r\right)$ repetitions for it to have a high chance of finding the
order whereas the mixed state Shor's algorithm uses exactly the same resources as
Shor's original algorithm but requires
\begin{equation}
\label{repmodshor}
    O\left(\frac{pq}{(p-1)(q-1)}\log \log r\right)
\end{equation}
repetitions for it to have a high chance of finding the order which, in the limit $p,q
\rightarrow \infty$, is equally as efficient as Shor's algorithm.
For simplicity we will prove this efficiency result %of equation \ref{repmodshor}
for a
mixed state algorithm with $L$ control qubits. For the reasons outlined above the result
will be identical using a single pure control qubit. The proof follows very closely that
of Shor \cite{Shor95}.

Pick an $L$ such the $N^2 < t=2^L < 2N^2$. The initial state of our system with all
the control qubits grouped into the first state is
\begin{equation}
\label{start}
\rho_{ini} = \frac{1}{Nt} \sum^{t-1}_{a=0} \sum^{t-1}_{b=0} \ket{a}\bra{b} \otimes
\sum_{d}
\sum^{r_d-1}_{j_d=0} \ket{\psi_j^d} \bra{\psi_j^d}.
\end{equation}
Application of the controlled $U_a, U_a^2,\cdots,U_a^{2^{L-1}}$ gates and the inverse
Fourier transform
yields the state
\begin{eqnarray}
\label{afterfourier}
\rho_2 = \frac{1}{Nt^2} \sum_d \sum^{r_d-1}_{j_d=0} \sum^{t-1}_{a,b,k,l=0}
e^{2 \pi i  a \left(\frac{j_d}{r_d} - \frac{k}{t} \right)}
\nonumber \\
e^{-2 \pi i  b \left(\frac{j_d}{r_d} - \frac{l}{t} \right)} \ket{k}
\bra{l} \otimes \ket{\psi_{j_d}^d} \bra{\psi_{j_d}^d}.
\end{eqnarray}

We now make a measurement on the first state.
The probability that the result $c$ is obtained is
\begin{eqnarray}
\label{prob}
P(c) =\frac{1}{Nt^2} \sum_{d} \sum^{r_d-1}_{j_d=0} \left| S \right| ^2, \, \,  S = \sum^{t-1}_{a=0}
e^{2 \pi i  a \left(\frac{j_d}{r_d} - \frac{c}{t} \right)}.
\end{eqnarray}
$S$ is just an arithmetic progression and $|S|^2$ can easily be bounded by
\begin{equation}
\label{modS}
|S|^2 > \frac{4t^2}{\pi^2} \qquad \hbox{for} \qquad \left| \frac{j_d}{r_{d}} -\frac{c}{t}
\right| < \frac{1}{2t}.
\end{equation}
Because $t>N^2$ this is a sufficient condition that given $c/t$ there is only
one fraction $j_d/r_d$ with $r_d<N$ such that the above condition is obeyed.
For a given measurement result $c$ there are at least $(p-1)(q-1)/r$ corresponding
values of $r_d$ with $r_d = r$ by theorem 1. So the probability that $c/t$
is the best estimate of a fraction with denominator $r$ is
\begin{equation}
\label{fbound}
P'(c) > \frac{1}{Nt^2} \sum_d \sum_{j_d=1}^{r_d-1} |S|^2 >  \frac{4(p-1)(q-1)}{N \pi^2
r}.
\end{equation}
We now require that the numerator, $j_d$, is coprime to $r$ otherwise
cancellation of common factors will occur in $j_d/r$. There are $\phi(r)$ values of $j_d$
which are less than and coprime to $r$, where $\phi$ is Euler's totient function \cite{Hardy}.
Thus the probability that we can calculate
$r$ is
$
%\begin{equation}
%\label{Pfind}
P > 4(p-1)(q-1)\phi(r)/N r \pi^2.
%\end{equation}
$
Using a theorem by Hardy and Wright (theorem 328) \cite{Hardy} that $\phi(r) /r >
\delta/
\log \log r$ for some constant $\delta$ we find that the number
of times that we need run the algorithm to have a high chance of finding the period, $r$
${}_{\Box}$is given by Eq. (\ref{repmodshor}).
%times
%\begin{equation}
%\label{finally}
%O\left(\frac{pq}{(p-1)(q-1)} \log \log r \right)
%\end{equation}

We have thus found that one pure qubit and a supply of maximally
mixed qubits is sufficient to implement Shor's algorithm,
requiring no more resources in terms of quantum operations or
physical systems than the algorithm operating on pure quantum states.
This implies that the algorithm presented here is a 'true' quantum
algorithm, achieving an exponential speedup using only polynomial
resources. This may be suprising as the degree of mixing of the state
of the computer is high. However, the mixing decrease as the algorithm proceeds but
never below a mixture of $N/r_d$ eigenstates where $r_d$ is the
{\it measured} period. Furthermore, it should be noted that despite this strong
degree of mixing the quantum computer actually evolves into an
entangled state. It is this entanglement that appears to be responsible for
the computational speedup.

Maximally mixed states are intuitively a less 'costly' resource than
pure states but, in fact, we do not need to require {\em maximally} mixed
states: we could equally well use
any random state (mixed or pure) on which to perform the controlled $U_a$
operations. The {\em average} efficiency over all these states would
then be as we have shown in this paper. In particular thermal states of
nuclear spins (e.g. in NMR), where the occupation of the ground state is only slightly
greater than that of the first excited state, would change the efficiency
of this algorithm by only a small amount leaving it an efficient algorithm.
This ability of highly mixed states to support efficient quantum computation
points towards the possibility of the implementation of true quantum computation
for example in NMR systems.

The authors would like to thank Kevin Buzzard for valuable advice on number theory
and S. Bose, R. Jozsa, R. Laflamme, M. Nielsen and C. Zalka for helpful comments.
This work is supported by the
%United Kingdom Engineering and Physical Sciences Research Council
EPSRC, the Leverhulme Trust,
%two EU TMR-networks ERB 4061PL95-1412 and ERB FMRXCT96-0066
and the EU project EQUIP.

\end{multicols}

\end{document}